\documentclass[10pt,conference]{IEEEtran}
\usepackage{cite}
\usepackage{amsmath,amssymb,amsfonts}
\usepackage{algorithmic}
\usepackage{graphicx}
\usepackage{textcomp}
\usepackage{xcolor}
\usepackage[hyphens]{url}
\usepackage{fancyhdr}
\usepackage{hyperref}
\usepackage{makecell}
\usepackage{multirow}
\usepackage{float}
\usepackage{verbatim}
\usepackage{booktabs}
\usepackage[inline]{enumitem}
\usepackage{caption}
\usepackage{subcaption}
\usepackage{amsmath}
\usepackage{MnSymbol}
\usepackage{wasysym}
\usepackage{tikz}
\usepackage{array}
\usepackage{tabularx}
\usetikzlibrary{tikzmark, matrix, backgrounds}
\usepackage{array}  
\usepackage{multicol}
\usepackage[ruled,vlined]{algorithm2e}

\newcommand{\hpcayear}{2026}

\newcommand{\name}{{\it DIAMOND}}
\newcolumntype{P}[1]{>{\centering\arraybackslash}p{#1}}

\definecolor{reviewercolor}{RGB}{0,0,150}
\definecolor{responsecolor}{RGB}{0,100,0}
\definecolor{highlight}{RGB}{220,0,0}

\begin{document}

\title{Systolic Array Acceleration of Diagonal-Optimized Sparse-Sparse Matrix Multiplication for Efficient Quantum Simulation}

\newcommand{\hpcapubid}{0000--0000/00\$00.00}
\newcommand\hpcaauthors{Yuchao Su, Srikar Chundury, Jiajia Li, Frank Mueller$\ddagger$}
\newcommand\hpcaaffiliation{North Carolina State University$\dagger$}
\newcommand\hpcaemail{\{ysu34, schundu3, jli256, fmuelle\}@ncsu.edu}



\author{
  \ifdefined\hpcacameraready
    \IEEEauthorblockN{\hpcaauthors{}}
      \IEEEauthorblockA{
        \hpcaaffiliation{} \\
        \hpcaemail{}
      }
  \else
    \IEEEauthorblockN{\hpcaauthors{}}
      \IEEEauthorblockA{
        \hpcaaffiliation{} \\
        \hpcaemail{}
      }
  \fi 
}

\fancypagestyle{camerareadyfirstpage}{%
  \fancyhead{}
  \renewcommand{\headrulewidth}{0pt}
  \fancyhead[C]{
    \ifdefined\aeopen
    \parbox[][12mm][t]{13.5cm}{\hpcayear{} IEEE International Symposium on High-Performance Computer Architecture (HPCA)}    
    \else
      \ifdefined\aereviewed
      \parbox[][12mm][t]{13.5cm}{\hpcayear{} IEEE International Symposium on High-Performance Computer Architecture (HPCA)}
      \else
      \ifdefined\aereproduced
      \parbox[][12mm][t]{13.5cm}{\hpcayear{} IEEE International Symposium on High-Performance Computer Architecture (HPCA)}
      \else
      \parbox[][0mm][t]{13.5cm}{\hpcayear{} IEEE International Symposium on High-Performance Computer Architecture (HPCA)}
    \fi 
    \fi 
    \fi 
    \ifdefined\aeopen 
      \includegraphics[width=12mm,height=12mm]{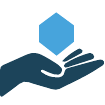}
    \fi 
    \ifdefined\aereviewed
      \includegraphics[width=12mm,height=12mm]{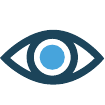}
    \fi 
    \ifdefined\aereproduced
      \includegraphics[width=12mm,height=12mm]{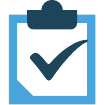}
    \fi
  }
  \fancyfoot[C]{}
}
\fancyhead{}
\renewcommand{\headrulewidth}{0pt}

\maketitle

\ifdefined\hpcacameraready 
  \thispagestyle{camerareadyfirstpage}
  \pagestyle{empty}
\else
  \thispagestyle{plain}
  \pagestyle{plain}
\fi

\newcommand{\hpcaheight}{0mm}
\ifdefined\eaopen
\renewcommand{\hpcaheight}{12mm}
\fi


\begin{abstract}
Hamiltonian simulation is a key workload in quantum computing,
enabling the study of complex quantum systems and serving as a
critical tool for classical verification of quantum devices. However,
it is computationally challenging because the Hilbert space
dimension grows exponentially with the number of qubits. The growing
dimensions make matrix exponentiation, the key kernel in Hamiltonian
simulations, increasingly
expensive. Matrix exponentiation is typically approximated by the Taylor series,
which contains a
series of matrix multiplications. Since Hermitian operators
are often sparse, sparse
matrix multiplication accelerators are essential for
improving the scalability of classical Hamiltonian simulation. Yet,
existing accelerators are primarily designed for machine learning
workloads and tuned to their characteristic sparsity patterns, which
differ fundamentally from those in Hamiltonian simulations that are
often dominated by structured diagonals.

In this work, we present \name, the first diagonal-optimized quantum
simulation accelerator. It exploits the diagonal structure commonly
found in problem-Hamiltonian (Hermitian) matrices and leverages a
restructured systolic array dataflow to transform diagonally sparse
matrices into dense computations, enabling high utilization and
performance. Through detailed cycle-level simulation of diverse
benchmarks in HamLib, \name{} demonstrates average performance
improvements of $10.26\times$, $33.58\times$, and $53.15\times$ over
SIGMA, Outer Product, and Gustavson's algorithm, respectively, with
peak speedups up to $127.03\times$ while reducing energy consumption
by an average of $471.55\times$ and up to $4630.58\times$ compared to
SIGMA.

\end{abstract}

\section{Introduction}
\label{sec:introduction}

Quantum computing has emerged as a promising computational paradigm
with the potential to accelerate solutions to classically intractable
problems. Notably, quantum algorithms have demonstrated theoretical
advantages for tasks such as combinatorial optimization, simulation of
quantum systems, and certain NP-complete problems like
Max-Cut~\cite{farhi2014quantum}, 3-SAT, and
beyond~\cite{montanaro2016quantum}. Industry efforts --- from IBM's
superconducting architectures~\cite{koch2007charge} and Google's
Sycamore processor~\cite{arute2019quantum} over IonQ's ion
traps~\cite{monroe2013scaling} to Microsoft's topological
approaches~\cite{freedman2003topological} --- have advanced quantum
hardware capabilities, but these systems remain limited by coherence
times, gate fidelity, and restricted quantum bit (qubit)
counts~\cite{preskill2018nisq}.
Architectural studies address these challenges via hardware‑aware
compilation~\cite{tannu2019asplos}, noise‑aware
mapping~\cite{murali2020isca}, surface‑code
adaptations~\cite{yin2024micro}, and inter‑chip
links~\cite{zhang2024asplos}.

Due to these limitations, classical simulation of quantum circuits
continues to play a central role in quantum algorithm development.
Simulators serve as essential tools for prototyping, debugging, and
benchmarking quantum applications, often long before they can run on
real hardware~\cite{qiskit, cirq_developers_2021_5182845}. Depending on
the target workload, simulators adopt different numerical strategies,
such as state vector evolution for general-purpose
circuits~\cite{haner2016high}, unitary
simulation~\cite{amy2019synthesis} for analyzing circuit matrices, or
Hamiltonian-based techniques for time
evolution~\cite{johansson2013qutip} and physics inspired
models~\cite{orus2019tensor}.

A key challenge in classical simulation is the exponential scaling of
the Hilbert space, which makes even modest quantum systems
computationally intensive.
This growth makes simulation a natural candidate for High-Performance Computing
(HPC), leveraging large-scale compute, memory, and bandwidth.
Recent HPC-oriented
simulators and accelerators~\cite{quest2018,wu2022isca} demonstrate the
importance of architectural support for scalable quantum simulation. To
mitigate the cost, recent efforts have begun to exploit sparsity in
quantum applications, either in the structure of quantum circuits or in
the operators they encode. For instance, Trotterized Hamiltonians and
domain-specific unitaries often yield matrices with block-diagonal or
sparse diagonal structure. Recognizing and leveraging these patterns
has enabled the development of custom numerical formats such as
DiaQ~\cite{diaq}, a more compact version of the traditional DIA
format~\cite{li2013smat}. While such sparse representations
significantly reduce memory footprint, they often increase
computational cost because arithmetic must now operate only on nonzeros
rather than on contiguous dense blocks. This shift amplifies the
importance of sparse matrix-matrix multiplication (SpMSpM), which
already dominates the runtime of many simulation methods and becomes
even more critical as sparsity grows. Whether simulating time evolution
or applying deep quantum circuits, the efficiency of the simulator
increasingly depends on high-throughput, sparsity-aware numerical
kernels that can handle extreme sparsity that often exceed 99\% (shown
in Table~\ref{tab:dataexample}, which is discussed detailed in Section~\ref{sec:eva})
while sustaining HPC-level performance.

Encoding Hermitian matrices in a diagonal format transforms extremely
sparse matrices into dense diagonal bands. This representation improves
data locality and reuse but requires a fundamental rethinking of
dataflow to be efficiently supported in hardware. Conventional systolic
arrays~\cite{kungsystolic}, though effective for dense matrix
multiplication, are ill-suited for such structured sparsity without
significant adaptation to avoid processing redundant elements. Hardware
microarchitectures that exploit predictable graph structures have been
proposed in quantum error correction
decoders~\cite{das2022hpca,vittal2023isca}, demonstrating how tailoring
dataflow to domain-specific patterns can deliver efficiency gains.
However, existing sparse matrix multiplication accelerators are
primarily optimized for machine learning
workloads~\cite{yao2025apspgemm,xie2019ia,bulucc2012parallel,nagasaka2018high,nagasaka2019performance},
and fail to exploit the diagonal regularity seen in quantum
applications. Prior diagonal-format accelerators~\cite{gao2021adaptive}
instead target Sparse Matrix Vector Multiplication (SpMV) using the
classic DIA format, where each diagonal must be padded to the same
length.
Padding reduces storage efficiency and wastes computation,
which worsens for SpMSpM due to diagonal interactions.
Therefore, a specialized architecture and diagonal-aware
dataflow are essential.

To address this need, we present \name, \underline{D}iagonal
\underline{I}nspired \underline{A}ccelerator for \underline{M}atrix
Multiplication \underline{O}n \underline{N}onzero
\underline{D}iagonals, to the best of our knowledge the
first diagonal-optimized quantum simulation accelerator that can be
dynamically adapted to execute sparse matrix times sparse matrix
(SpMSpM) multiplication predominant in Hamiltonian simulation.
\name~features:
\begin{enumerate*}[label=(\alph*)]

\item a novel Diagonal Processing Element (DPE) that ensures data
  correctness,
  
\item a diagonal accumulator that supports the accumulation of partial
  results along diagonals,

\item a systolic array-like topology that exploits the inner grid data
  reuse pattern, and

\item a two-level memory hierarchy with

\item an efficient blocking strategy.

\end{enumerate*}
The complete \name~system connects multiple DPEs via a lightweight
global network-on-chip (NoC). Inside the NoC, each diagonal is
associated with a dedicated accumulator to collect results. \name~thus
effectively morphs extremely sparse matrix multiplication into dense
diagonal matrix multiplication.

\textit{Our key contributions are as follows:}
\begin{enumerate}

\item Analysis of diagonal matrix multiplication to motivate
  acceleration of diagonal sparse matrices,

\item a novel accelerator architecture, \name, for efficient diagonal
  sparse SpMSpM operations,

\item a diagonal accumulator design for efficient partial sum
  accumulation in diagonal format,

\item an efficient blocking strategy that enhances data locality and
  minimizes memory overhead, and

\item experimental results that show average speedups of $10.26\times$,
  $33.58\times$, and $53.15\times$ over SIGMA~\cite{SIGMA}, Outer
  Product, and Gustavson (with the latter two 
  originating from Flexagon~\cite{flexagon}), respectively, along with
  a $471.55\times$ average energy reduction compared to SIGMA.
\end{enumerate}

The remainder of this paper is organized as follows:
Section~\ref{sec:background} discusses modern Hamiltonian simulation
and diagonal-oriented storage formats.  Section~\ref{sec:math} details
the mathematical formulation for diagonal-based matrix multiplication.
Section~\ref{sec:design} presents the microarchitecture of \name{}.
Section~\ref{sec:eva} describes the benchmarks, experimental setup,
and performance evaluation against state-of-the-art accelerators.
Section~\ref{sec:related_work} reviews related work, and
Section~\ref{sec:conclusion} concludes the paper.

\section{Background}
\label{sec:background}

\subsection{Hamiltonian Simulations}
\label{subsec:quantum}
Quantum simulation workloads, such as previously introduced in
Section~\ref{sec:introduction}, span a diverse range of numerical
techniques ranging from full state vector evolution to structured
Hamiltonian dynamics.
Among these, simulations with sparse time-evolution operators are particularly
challenging due to their extreme size and sparsity.
These matrices are
often over 99\% sparse and, unlike unstructured ML workloads, exhibit
highly regular patterns aligned along fixed diagonals due to the
underlying physics.

\emph{Hamiltonian simulation}~\cite{ham} focuses on simulating the time evolution operator:
\begin{equation}
    U(t) = e^{-i H t},
    \label{eq:exponentiation}
\end{equation}
where \(H\) is a Hermitian Hamiltonian. The evolved quantum state is
obtained by applying this operator to the initial state vector:
\begin{equation}
    \psi(t) = U(t) \, \psi(0) = e^{-i H t} \, \psi(0).
    \label{eq:state_evolution}
\end{equation}
Directly computing the matrix exponential in
Eq.~\eqref{eq:exponentiation} is prohibitively expensive for large
\(N = 2^n\)-dimensional matrices, as it entails dense general-purpose matrix multiplication (GEMM) \(O(N^3)\) operations. To make this computation
tractable, quantum
simulation frameworks employ approximations such as \emph{Trotterization}
or the truncated \emph{Taylor series} expansion, which decompose the
evolution into products of sparse factors:
\begin{equation}
    e^{A} \approx \sum_{k=0}^{K} \frac{A^k}{k!}.
    \label{eq:taylors_expansion}
\end{equation}
Applying the Taylor series to a small timestep \( t/K \), the full
evolution \( e^{-iHt} \) is approximated as a product of \( K \)
short-time expansions:
\begin{equation}
    e^{-i H t} \approx 
    \left( \sum_{k=0}^{K} 
    \frac{(-i H \, t / K)^k}{k!} \right)^{K}.
    \label{eq:hamiltonian_simulation}
\end{equation}
Each short-time expansion involves matrix exponentiation and
can be decomposed into a series of SpMSpM operations (with one
additional SpMSpM per iteration).
Hence, SpMSpM is a dominant computational kernel in Hamiltonian simulations,
particularly in scenarios
involving time evolution via Trotterization.
To efficiently multiply them, data reuse
patterns and storage formats are required that differ significantly
from conventional assumptions.

\subsection{Quantum-Tailored Diagonal Storage Format}
\label{subsec:diaq}
Matrices pertinent to Hamiltonian simulations frequently exhibit
structured diagonal sparsity, with non-zero values
confined to a small number of fixed diagonal offsets. General-purpose
formats, such as CSR, CSC, and COO, represent each non-zero entry with
explicit row and column indices, incurring metadata overhead and
leading to irregular memory access. For example, CSR-based sparse
matrix-vector multiplication (SpMV) in libraries like SciPy operates
in $O(N + \text{nnz})$ time, where $N = 2^n$ is the state vector size
and $\text{nnz}$ is the number of non-zeros~\cite{scipy2020}.

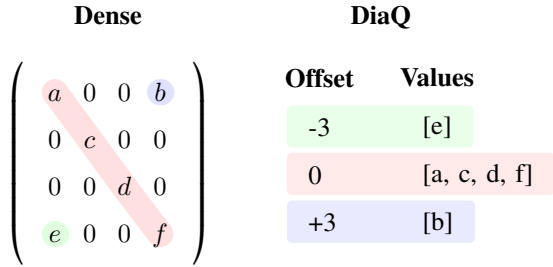
\begin{figure}[htb]
\centering
\renewcommand{\arraystretch}{1.1}
\begin{tabular}{@{}c@{}c@{}}
\textbf{Dense} & \textbf{DiaQ} \\
\begin{tikzpicture}[baseline=(current bounding box.north)]
  \matrix[matrix of math nodes,
          nodes={inner sep=4pt, text height=1.8ex, text depth=0.4ex, font=\normalsize},
          left delimiter={(}, right delimiter={)}] (m) {
    \node[alias=d0a]{a}; & 0 & 0 & \node[alias=d3]{b}; \\
    0 & \node[alias=d0b]{c}; & 0 & 0 \\
    0 & 0 & \node[alias=d0c]{d}; & 0 \\
    \node[alias=dm3]{e}; & 0 & 0 & \node[alias=d0d]{f}; \\
  };
  \begin{scope}[on background layer]
    \draw[red!30, line width=3.5mm, line cap=round, opacity=0.4]
      (d0a.center) -- (d0b.center) -- (d0c.center) -- (d0d.center);
    \draw[blue!30, line width=3.5mm, line cap=round, opacity=0.4]
      (d3.center) -- (d3.center);
    \draw[green!30, line width=3.5mm, line cap=round, opacity=0.4]
      (dm3.center) -- (dm3.center);
  \end{scope}
\end{tikzpicture}
&
{\hskip 8mm}
\begin{tikzpicture}[baseline=(current bounding box.north)]
  \matrix[column sep=10mm, row sep=2mm] (diaq) {
    \node[inner sep=2pt, font=\normalsize, anchor=west] (off1) {-3}; &
    \node[inner sep=2pt, font=\normalsize, anchor=west] (val1) {[e]}; \\
    \node[inner sep=2pt, font=\normalsize, anchor=west] (off2) {0}; &
    \node[inner sep=2pt, font=\normalsize, anchor=west] (val2) {[a, c, d, f]}; \\
    \node[inner sep=2pt, font=\normalsize, anchor=west] (off3) {+3}; &
    \node[inner sep=2pt, font=\normalsize, anchor=west] (val3) {[b]}; \\
  };
  \node[font=\bfseries\normalsize, anchor=south] at ([yshift=2mm]off1.north) {Offset};
  \node[font=\bfseries\normalsize, anchor=south] at ([yshift=2mm]val1.north) {Values};
  \begin{scope}[on background layer]
    \fill[green!20, opacity=0.4, rounded corners=2pt]
      ([xshift=-6pt,yshift=-2pt]off1.south west)
      rectangle
      ([xshift=6pt,yshift=2pt]val1.north east);

    \fill[red!20, opacity=0.4, rounded corners=2pt]
      ([xshift=-6pt,yshift=-2pt]off2.south west)
      rectangle
      ([xshift=6pt,yshift=2pt]val2.north east);

    \fill[blue!20, opacity=0.4, rounded corners=2pt]
      ([xshift=-6pt,yshift=-2pt]off3.south west)
      rectangle
      ([xshift=6pt,yshift=2pt]val3.north east);
  \end{scope}
\end{tikzpicture}
\end{tabular}
\caption{A diagonal matrix in dense format vs. DiaQ format. Each diagonal in
DiaQ can have a different length (deduced from the matrix 
dimensions and the offset), which eliminates the need for padding and saves 
space compared to the traditional DIA format.}
\label{fig:dense_vs_diaq}
\end{figure}

In contrast, DiaQ~\cite{diaq} stores sparse quantum operators as
collections of dense diagonals indexed by offset, eliminating
per-entry metadata and enabling memory-aligned access, as seen in
Fig.~\ref{fig:dense_vs_diaq}. Their format differs from the
traditional DIA format in that it does not require storing placeholder
\texttt{NA} values, though this comes at the cost of reduced memory
contiguity.  By skipping identity regions and directly traversing
active diagonals, it achieves $O(dN)$ complexity with fused SIMD
kernels, where $d$ is the number of relevant diagonals (typically
constant) yielding improved arithmetic intensity and locality.
Its design choice to skip \texttt{NA} values makes it particularly
effective for matrices exhibiting extreme sparsity, where diagonals
can be exponentially far apart. Such cases are frequently encountered
in sparse problem Hamiltonians (Hermitian matrices), where the shorter
diagonals enable more efficient storage than the standard DIA
format. Such characteristics make their storage layout particularly
effective when applying unitary gates of the form
$(I \otimes G \otimes I)$ to quantum states, where $G$ is a small
matrix acting on a subset of qubits.

While this quantum-tailored sparse format was previously integrated
with SV-Sim~\cite{svsim}, a state vector simulator, it also supports
structured SpMSpM via diagonal-wise products, which form the basis for
Hamiltonian simulations, as discussed in
subsection~\ref{subsec:quantum}. Instead of relying on dense $O(N^3)$
multiplication or general-purpose sparse kernels, it performs
$O(d^2N)$ compositions when each operand has $d$ diagonals. This
design benefits from sequential memory access, low indexing overhead,
and tight loop nests suited to modern vector units. Hence, domain-specific
diagonal formats are a natural fit for quantum
workloads where sparsity is aligned and predictable. They expose
regularity absent in traditional formats, making them well suited for
HPC accelerators targeting large-scale quantum simulations.
This focus on regularity aligns with broader architectural efforts that co-design
hardware and structured workloads~\cite{yin2024micro}.

However, conventional sparse accelerators are not designed for
diagonal-banded formats and fail to fully exploit the reuse and
regularity these formats expose. Unlocking their full potential
requires hardware that natively understands and accelerates diagonal
operations. This motivates the design of \name{}, a specialized
accelerator architecture that leverages diagonal structure to enable
efficient sparse matrix-matrix multiplication for quantum simulation
workloads.

\section{Diagonally Sparse Matrix Multiplication}
\label{sec:math}

We reformulate matrix multiplication into diagonal-wise computation
for efficient acceleration that leverages the diagonal sparse format.

\subsection{Diagonal Sparse Matrix Decomposition}

\begin{figure}
    \centering
    \includegraphics[width=0.7\linewidth]{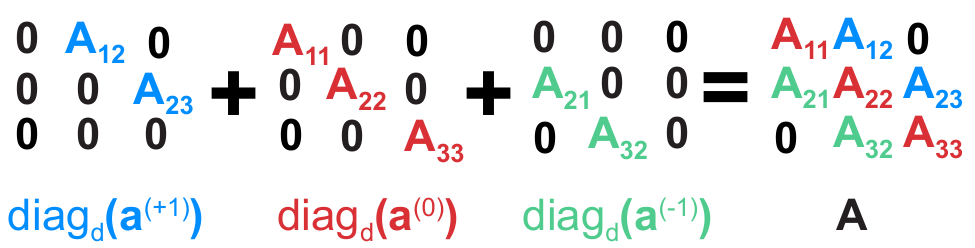}
    \caption{Diagonal Sparse Matrix Decomposition.}
    \label{fig:split}
    \vspace{-0.2in}
\end{figure}
To exploit diagonal sparsity, we recast matrix multiplication entirely
in diagonal space. Let $A, B \in \mathbb{R}^{N \times N}$. Recall that
each matrix can be uniquely expressed as the sum of its diagonals
\begin{equation}
A = \sum_{d \in D_A} \mathrm{diag}_{d}\bigl(\mathbf{a}^{(d)}\bigr),
\quad
B = \sum_{d \in D_B} \mathrm{diag}_{d}\bigl(\mathbf{b}^{(d)}\bigr),
\label{eq:diag_decomposition}
\end{equation}
where
\begin{itemize}
\item $D_A$ and $D_B$ are the sets of diagonal offsets containing
  nonzeros, e.g., $d=0$ is the main/principle diagonal, $d=1$ is the
  first (upper) superdiagonal, $d=-1$ the first (lower) subdiagonal;
\item $\mathbf{a}^{(d)}$ is the vector of entries along diagonal $d$
  in matrix $A$, often stored contiguously in diagonal formats; and
\item $\mathrm{diag}_{d}(\mathbf{v})$ denotes the operator that
  creates a matrix whose diagonal $d$ contains vector $\mathbf{v}$,
  such that all other entries are zero.
\end{itemize}

Formally, for any offset $d$, this operator is defined by
\begin{equation}
\bigl[\mathrm{diag}_{d}(\mathbf{v})\bigr]_{i,j} =
\begin{cases}
v_i, & j - i = d,\\
0, & \text{otherwise}.
\end{cases}
\label{eq:diag_operator}
\end{equation}

This formulation allows us to express $A$ and $B$ as structured sums
over their diagonals, as shown in Fig.~\ref{fig:split}.

\subsection{Offset Additivity}

A key property of diagonal multiplication is that multiplying two
diagonals yields a new diagonal whose offset is the sum of the input
offsets
\begin{equation}
\mathrm{diag}_{\,d_A + d_B}(\cdot)\;:=\;\ \mathrm{diag}_{d_A}(\cdot)\,\mathrm{diag}_{d_B}(\cdot).
\label{eq:offset_additivity}
\end{equation}

\noindent
\textbf{Proposition (Offset-Sum Rule).}
Let $d_A \in D_A$ and $d_B \in D_B$. Then any partial product between
these diagonals contributes exclusively to the result diagonal,
described as $d_C = d_A + d_B.$

\subsection{Diagonal Convolution Form}

Using Eq.~\eqref{eq:offset_additivity}, the matrix product $C = A B$ can
be rewritten as a sum over output diagonals:
\begin{equation}
C = \sum_{d_C \in D_C}
    \mathrm{diag}_{d_C}\Bigl(
        \sum_{\substack{d_A \in D_A,\, d_B \in D_B \\ d_A + d_B = d_C}}
        \mathbf{a}^{(d_A)} \odot \mathbf{b}^{(d_B)}
    \Bigr),
\label{eq:diag_convolution}
\end{equation}
where $\odot$ denotes element-wise multiplication over the overlapping
index range of $\mathbf{a}^{(d_A)}$ and
$\mathbf{b}^{(d_B)}$. Specifically, since diagonals
$\mathbf{a}^{(d_A)}$ and $\mathbf{b}^{(d_B)}$ may have different
lengths depending on their offsets, their element-wise product is
computed only over the range where both vectors are defined, i.e., at
the intersection of row and column indices. $D_C$ is defined as
\begin{equation}
D_C = D_A \oplus D_B = \bigl\{\, d_A + d_B \mid d_A \in D_A,\, d_B \in D_B \bigr\},
\label{eq:minkowski_sum}
\end{equation}
the Minkowski sum of the offset sets.


The diagonal-space reformulation expresses matrix multiplication as a
diagonal-wise convolution using the offset-sum rule. This isolates
nonzero interactions, eliminates redundant computation on zeros, and
exposes a natural structure for parallel and memory-efficient
execution. It forms the mathematical basis for our sparsity-aware
acceleration strategy.

\section{\name~Architecture}
\label{sec:design}

\begin{figure}[b]
    \centering
    \vspace*{-0.2in}
    \includegraphics[width=0.7\linewidth]{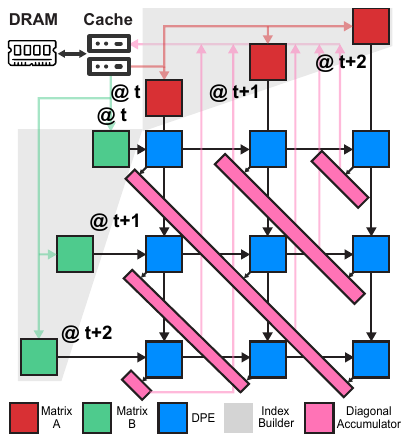}
    \caption{Overview of \name.}
    \label{fig:overview}
\end{figure}

\name{} comprises a grid of Diagonal Processing Elements (DPEs),
multiple diagonal accumulators, DRAM, and a set-associative
cache. Fig.~\ref{fig:overview} illustrates the high-level
architecture. DPEs, detailed in Section~\ref{subsec:DPE}, are
interconnected via a lightweight Network-on-Chip (NoC) to form a
scalable compute fabric. Input diagonals from matrix~$A$ are streamed
from the top in increasing offset order, and those from matrix~$B$
from the left in decreasing order. Prior to entering the grid, a
lightweight index builder extracts the offset from the diagonal format to initialize element indices efficiently. Each diagonal enters at cycle
$t, t+1, t+2, \dots$, enabling a synchronized dataflow. The DPE grid
is dynamically sized: the number of columns equals the number of
active diagonals in $A$ and rows in $B$. The anti-diagonal
accumulation naturally aligns with the Minkowski-sum-based dataflow
(Section~\ref{sec:math}).

Unlike MAC-based accelerators~\cite{tpu,eyeriss}, \name{} assigns each
output diagonal to a dedicated accumulator
(Section~\ref{subsec:DACC}). A set-associative cache improves
efficiency by capturing temporal locality in diagonal reuse
(Section~\ref{subsec:cache}). Data is fetched from DRAM into cache,
then fed into the grid. Each diagonal occupies a full column or row,
and only the first element's position is explicitly encoded; all
others are derived by self-increment, as shown by the gray
blocks in Fig.~\ref{fig:overview}.

To scale to larger matrices, \name{} applies a two-dimensional
diagonal blocking strategy (Section~\ref{subsec:block}), which bounds
the grid size and enhances memory locality. Section~\ref{subsec:cycle}
details the execution timeline (preload, compute, and popout) and
provides a cycle-level performance model for sparse workloads. A
complete example of data movement and computation is presented in
Section~\ref{subsec:example}.


\subsection{Microarchitecture of the DPE}
\label{subsec:DPE}

\begin{figure}[t]
    \centering
    \includegraphics[width=0.8\linewidth]{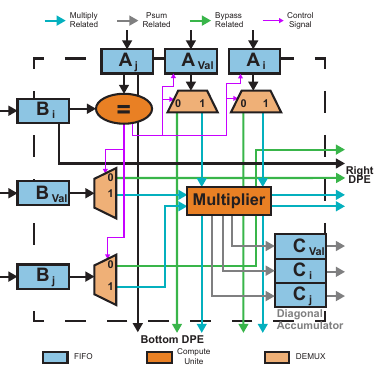}
    \caption{Microarchitecture of DPE.}
    \label{fig:dpe}
    \vspace*{-0.2in}
\end{figure}

Each DPE consists of a comparator, a multiplier, four DEMUXes, six
input FIFOs (three from the left and three from the top), and three
output FIFOs (see Fig.~\ref{fig:dpe}). All FIFOs are of size one and
are used to store the row index ($i$), column index ($j$), and value
of each operand. The left-side FIFOs buffer data from matrix~$B$,
while the top-side FIFOs buffer data from $A$.

A distinguishing feature of the DPE is its comparator, which enables
dynamic alignment of diagonals, a mechanism inspired by the control
logic used in SpMV accelerators~\cite{sparseTPU}. Unlike conventional
systolic arrays, where processing elements operate under fixed
neighboring relationships, 
\name{} must accommodate arbitrary and irregular diagonal patterns
without prior structural assumptions. The comparator in each DPE
evaluates whether the column index of the matrix~$A$ element matches
the row index of the $B$ element.  If a match is detected (see
Table~\ref{tab:comparator}), the operands are sent to the multiplier
to compute a partial product, which is then forwarded to the output
FIFOs and subsequently to the corresponding diagonal accumulator.

\begin{table}[htb]
\centering
\caption{Comparator Operating Logic.}
\label{tab:comparator}
\begin{tabular}{@{}ccc@{}}
\toprule
Condition & Action  \\
\midrule
$j_A = i_B$   & Forward to Multiplier \& Multiply \\
$j_A \neq i_B$ & Stall the data with larger index \\
Missing one   & Forward existing data\\
Missing both  & Wait for valid data \\
\bottomrule
\end{tabular}
\end{table}

The detailed comparison logic is as follows:
\begin{itemize}
\item If a match is detected, the two values are multiplied.
\item If the indices differ, the DPE retains the data with the larger
  index and forwards the other operand to its neighboring DPE
  ($A$ data downward and $B$ data rightward).
\item If the DPE receives only one operand in a cycle (from either $A$
  or $B$), it forward the operand to its neighbor PE(A to bottom and B to right).
\end{itemize}
This strategy builds on the observation that the indices of nonzero
elements along each diagonal increase monotonically as data flows
through the systolic array. The original matrix coordinates $(i, j)$
can be determined from the diagonal format with the diagonal
index. The coordinate calculation is performed only for the first
element of each diagonal; subsequent positions are obtained through
self-increment, eliminating the repeated index computation.

When two operands arrive at a DPE --- one from matrix~$A$ with index
pair $(i, j_A)$ and one from $B$ with index pair $(i_B, j)$ --- the
comparator checks if the inner dimensions align, i.e., if $j_A =
i_B$ is true, a valid multiplication is possible.

Otherwise, assuming $j_A < i_B$, the operand from $A$ is considered
``behind'' the operand from $B$. Since the column indices from $A$ and
the row indices from $B$ both increase in future cycles, any upcoming
$A$ elements in this row will have $j > j_A$, moving even further away
from $i_B$. Thus, the current A operand can no longer match any future
B operands in this column and is forwarded downward, while the B
operand is retained.

Conversely, if $j_A > i_B$, then the $B$ operand is behind and is
forwarded rightward, while the $A$ operand is held. This selective
forwarding ensures that operands are only retained when a future match
remains possible, thereby avoiding unnecessary computations and
optimizing data movement within the array.

The above DPE design is critical to \name{}, while the comparator 
plays a pivotal role. By integrating index matching and value holding 
within each DPE, \name{} can effectively handle the irregular structure
of diagonal sparse matrices and ensure the correctness of partial
sums.
\subsection{Diagonal Accumulator}
\label{subsec:DACC}

The structure of \name's dataflow is derived from the Minkowski-sum
formulation described in Section~\ref{sec:math}, which governs how
diagonals from the input matrices contribute to diagonals in the
output matrix. Specifically, each DPE is
mapped to a unique pair of input diagonals from $A$ and $B$, where $A$
is streamed into the array from the top and $B$ is streamed from the
left. Let $d_i$ and $d_j$ denote the diagonal offsets fed into row~$i$
and column~$j$, respectively. The DPE at position $(i,j)$ is
responsible for computing partial sums that contribute to the output
diagonal at offset $d_C = d_i + d_j$, as determined by the
Minkowski-sum rule.

\begin{figure}[htb]
     \centering
     \begin{subfigure}[b]{0.4\linewidth}
         \centering
         \includegraphics[width=0.9\textwidth]{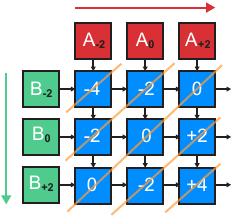}
         \caption{}
         \label{fig:anti1}
         \vspace{0.1in}
     \end{subfigure}
     \hspace{0.02\linewidth}
     \begin{subfigure}[b]{0.4\linewidth}
         \centering
         \includegraphics[width=0.9\textwidth]{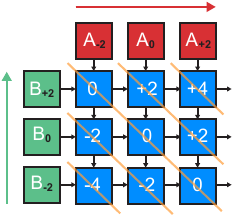}
         \caption{}
         \label{fig:dia1}
         \vspace{0.1in}
     \end{subfigure}
     \begin{subfigure}[b]{0.4\linewidth}
         \centering
         \includegraphics[width=0.9\textwidth]{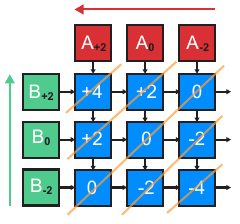}
         \caption{}
         \label{fig:anti2}
     \end{subfigure}
    \hspace{0.02\linewidth}
    \begin{subfigure}[b]{0.4\linewidth}
         \centering
         \includegraphics[width=0.9\textwidth]{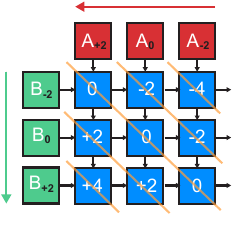}
         \caption{}
         \label{fig:dia2}
     \end{subfigure}
         \caption{Minkowski-Sum Mapping.}
        \label{fig:mapping}
        \vspace*{-0.1in}
\end{figure}

This direct mapping allows us to examine how data flows through the
DPE grid and reveals patterns in the spatial alignment of computation
and accumulation. Fig.~\ref{fig:mapping} illustrates four possible
feeding strategies and the resulting accumulation patterns:

\begin{itemize}
\item Fig.~\ref{fig:anti1} shows both $A$ and $B$ fed in
  increasing order of diagonal offset. In this case, all DPEs along
  the same anti-diagonal contribute to the same output
  diagonal.
\item Fig.~\ref{fig:dia1} feeds $A$ in increasing order and
  $B$ in decreasing order, producing a diagonal pattern in the grid,
  where each diagonal contributes to the same output diagonal.
\item Fig.~\ref{fig:anti2} feeds both $A$ and $B$ in
  decreasing order. This also results in an
  anti-diagonal contribution pattern.
\item Fig.~\ref{fig:dia2} feeds $A$ in decreasing order and
  $B$ in increasing order, again forming a diagonal contribution
  pattern.
\end{itemize}

Across all configurations, a consistent structural property emerges:
DPEs aligned along a given diagonal or anti-diagonal in the compute
grid contribute to the same output diagonal. This observation
naturally leads to the use of diagonal accumulators in \name. Each
accumulator is assigned to a unique output diagonal and gathers
partial sums from all corresponding DPEs. Since output diagonals are
mutually independent, the accumulation process is highly
parallelizable.

The accumulators are implemented using conventional hardware, i.e.,
they do not require specialized logic beyond index decoding. Their
structure is simple but well-aligned with the deterministic dataflow
exposed by the Minkowski-based mapping, enabling both efficiency and
scalability in accumulation.

\subsection{Blocking}
\label{subsec:block}



Matrix sizes in Hamiltonian simulations grow exponentially. For
example, a 10-qubit gate yields a $1024 \times 1024$ matrix, with
dimensions scaling as $2^n$ for $n$ qubits. Although diagonals formats
often omit entirely zero diagonals, the length of non-zero diagonals
remains tied to the full matrix size, given by
$\mathrm{L}_{D_i} = N - |i|$, where $N$ is the matrix dimension and
$i$ is the diagonal index. These long diagonals demand large buffers.

\begin{figure}[htb]
    \centering
    \includegraphics[width=0.7\linewidth]{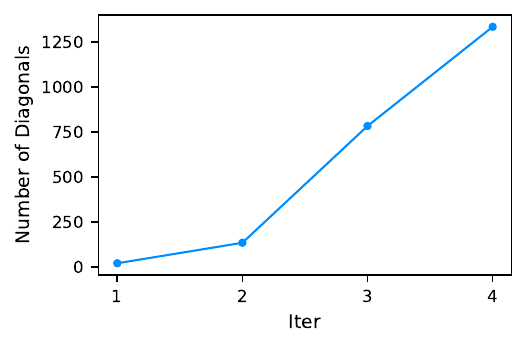}
        \vspace{-0.3cm}
        \caption{\textbf{Rapid growth} of the number of
          \emph{non-zero} diagonals during 10-qubit Heisenberg
          Hamiltonian simulation. Each iter corresponds to a step in
          the chained matrix multiplication.}
    \label{fig:diagonalnum}
    \vspace*{-0.2in}
\end{figure}

Moreover, as chained multiplications proceed, the number of non-zero
diagonals increases (see Fig.~\ref{fig:diagonalnum}). Without
blocking, this growth would require dynamic scaling of the DPE
grid. To maintain a fixed and feasible grid size, efficient blocking
becomes critical. We identify two key strategies: row/column-wise
blocking and diagonal blocking.

\subsubsection{Row/Col-wise Blocking}

To address the problem of long diagonals, we partition the diagonals
based on rows and columns. Since matrix multiplication involves
multiplying rows of $A$ with columns of $B$, we partition $A$
column-wise and $B$ row-wise. This restricts the column range in $A$
and the row range in $B$.

As long as both matrices are partitioned at the same row/column
indices, this strategy generates aligned group pairs. Each column
group of $A$ has a one-to-one correspondence with a row group of $B$
because they share the same index range. Groups with non-overlapping
index ranges do not need to be multiplied, as their respective row and
column indices do not intersect, which means there would be no valid
$(i,j,k)$ triplets satisfying the multiplication condition
$A_{i,k} \cdot B_{k,j}$. In other words, if the column indices of $A$
in one group do not match the row indices of $B$ in the corresponding
group, their product contributes nothing to the result
matrix. Therefore, multiplying mismatched groups is unnecessary and
wasteful.

\begin{figure}[b]
    \vspace*{-0.2in}
     \centering
     \begin{subfigure}[b]{0.4\columnwidth}
         \centering
         \includegraphics[width=0.8\textwidth]{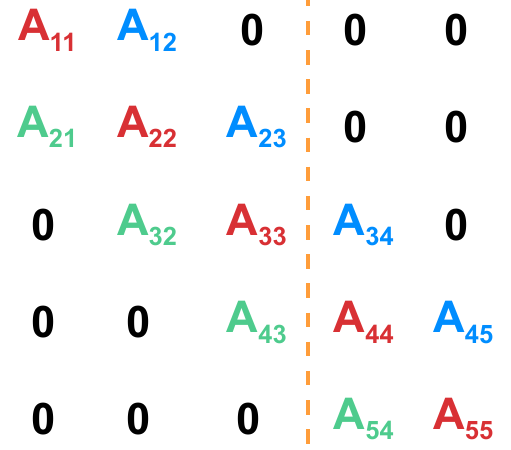}
         \caption{Column Groups}
         \label{fig:colgroup}
     \end{subfigure}
     \hspace{0.01\linewidth}
     \begin{subfigure}[b]{0.4\columnwidth}
         \centering
         \includegraphics[width=0.8\textwidth]{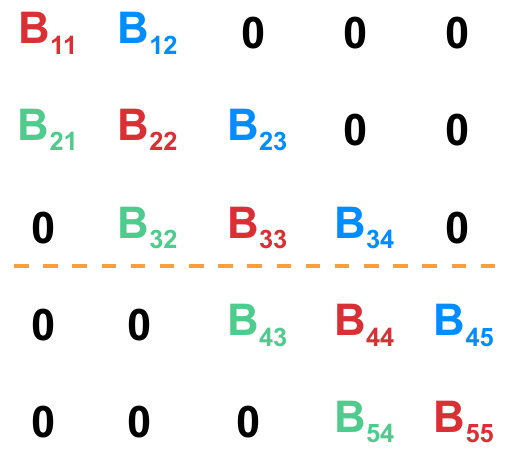}
         \caption{Row Groups}
         \label{fig:rowgroup}
    \end{subfigure}
    \hfill
     \begin{subfigure}[b]{0.4\columnwidth}
         \centering
         \includegraphics[width=0.8\textwidth]{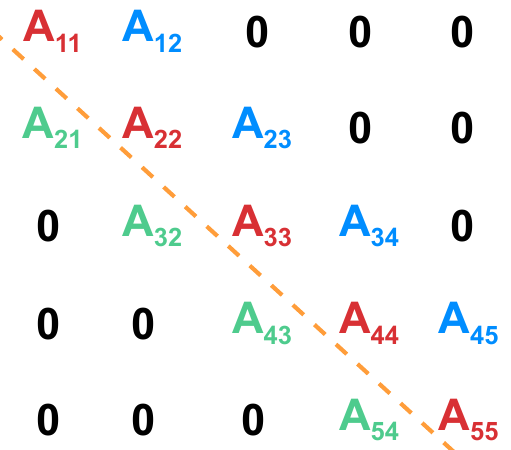}
         \caption{Diagonal Groups}
         \label{fig:diagroup1}
     \end{subfigure}
     \hspace{0.01\linewidth}
     \begin{subfigure}[b]{0.4\columnwidth}
         \centering
         \includegraphics[width=0.8\textwidth]{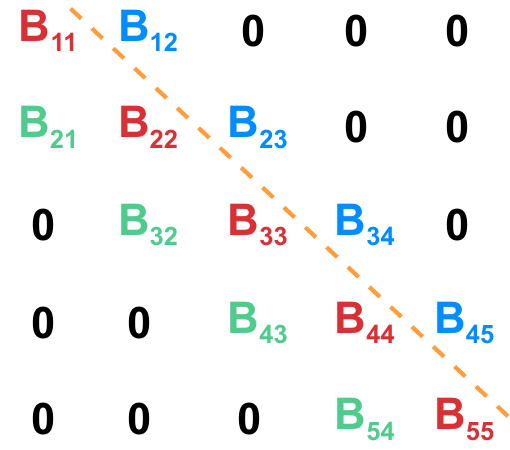}
         \caption{Diagonal Groups}
         \label{fig:diagroup2}
     
     \end{subfigure}
         \caption{Blocking Strategies.}
        \label{fig:colrowgroup}
    \vspace*{-0.1in}
\end{figure}

As shown in Fig.~\ref{fig:colgroup}, $A$ is partitioned at column
index 3 (1-based), yielding two groups: columns 1--3 and
4--5. Simultaneously, $B$ is row-partitioned at the same index
(Fig.~\ref{fig:rowgroup}), forming two matching groups: rows 1--3 and
4--5. This reduces a problem with diagonals up to length 5 into two
smaller subproblems with maximum diagonal lengths of 3 and 2.

The same method can be applied to much larger matrices by creating
more groups. Importantly, only matching group pairs from the two
matrices need to be multiplied as cross-group products are irrelevant
due to index mismatch. This technique significantly reduces cache-line
overhead by limiting the number of elements stored per line, thereby
improving cache efficiency.

\subsubsection{Diagonal Blocking}

Although the matrix remains sparse during the Hamiltonian simulation, the
number of diagonals grows exponentially, as shown in
Fig.~\ref{fig:diagonalnum}, making the DPE grid size difficult to
manage. To address this, a diagonal blocking algorithm is necessary.

In classical dense matrix multiplication, all rows in $A$ must be
multiplied with all columns in $B$. Similarly, as described in
Section~\ref{sec:math}, diagonal matrix multiplication requires each
diagonal in $A$ to be multiplied with every diagonal in $B$. This
implies that each pairwise multiplication between diagonals is
independent of the others.

Given this property, unlike row/column-wise blocking, we can partition
the diagonals of $A$ and $B$ independently into different numbers of
groups. This flexibility is especially useful in quantum simulation
workloads. For instance, in Hamiltonian simulation, $A$ (resulting
from the previous multiplication) contains a growing number of
diagonals, whereas $B$ consistently maintains the same number of
diagonals. In such a scenario, $A$ can be divided into multiple
diagonal groups, while $B$ remains as a single group.

After grouping, each diagonal group in $A$ must be multiplied with all
groups in $B$. As shown in
Figs.~\ref{fig:diagroup1},~\ref{fig:diagroup2}, partition boundaries
do not need to align. For example, $A$ is split at diagonal $-1$ and
$B$ at $0$, each forming two groups. Each diagonal in $A$ maps to a
grid column and each in $B$ to a row. Thus, the DPE grid size is
determined by the maximum number of diagonals in any group. In this
case, a $2 \times 2$ grid suffices.

The full multiplication comprises four group-wise computations:
$(A_1, B_1)$, $(A_1, B_2)$, $(A_2, B_1)$, and $(A_2, B_2)$. Each pair
utilizes the grid, though some rows or columns may remain idle when
the group has fewer diagonals. Execution order is flexible and does
not affect correctness.

Although the example is simple, diagonal blocking is critical for
scalability. For instance, a 10-qubit Heisenberg Hamiltonian
simulation yields 783 diagonals in the third iteration. Blocking these
into groups of 64 or 256 limits grid dimensions and reduces hardware
demands. Additionally, such blocking improves temporal locality, as
all multiplications within a group pair reuse loaded data before
switching to the next pair.

\subsection{Memory System and Data Locality}
\label{subsec:cache}

To evaluate the impact of data locality in the \name~dataflow, we
design a simple two-level memory system. Our goal is not to simulate
detailed memory behavior, but rather to observe how the blocking
algorithm affects memory access patterns and footprint. This
abstraction allows us to measure the effectiveness of data locality,
rather than modeling real-world overheads.

\subsubsection{Memory System}
The memory system consists of two layers: a set-associative cache and
a backing DRAM. The cache connects to the left and top inputs of the
DPE grid and to the diagonal accumulators. All DRAM accesses are
mediated by the cache, enabling simulation of latency differences
between fast and slow memory.

Each cache line holds one diagonal block group, and a Least Recently
Used (LRU) policy is applied. Cache hits incur 1 cycle, while misses
add a 5-cycle LRU penalty and trigger a DRAM access. DRAM reads and
writes incur a fixed 50-cycle latency, modeling the cost gap between
local and remote memory and highlighting the importance of
locality-aware blocking.

\subsubsection{Intra-Grid Locality}

\begin{figure}[htb]
    \centering
    \vspace*{-0.1in}
    \begin{subfigure}[t]{0.4\columnwidth}
        \centering
        \includegraphics[width=0.7\linewidth]{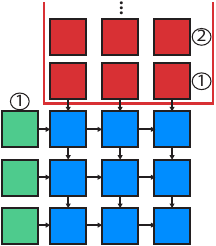}
        \caption{Inter-Block Locality}
        \label{fig:outerreuse}
    \end{subfigure}
    \hfill
    \begin{subfigure}[t]{0.48\columnwidth}
        \centering
        \includegraphics[width=0.3\linewidth]{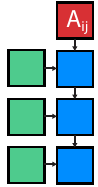}
        \caption{Intra-Grid Locality(Column)}
        \label{fig:inner1}
    \end{subfigure}

    \vskip\baselineskip
    \begin{subfigure}[b]{0.7\linewidth}
        \centering
        \includegraphics[width=0.4\linewidth]{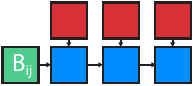}
        \caption{Intra-Grid Locality(Row)}
        \label{fig:inner2}
    \end{subfigure}

    \caption{Reuse patterns in the \name~dataflow.}
    \label{fig:reuse_patterns}

\end{figure}

\name~exhibits two distinct types of data locality. The first is
\textit{intra-grid locality}, enabled by the systolic array
structure. Once data enters the DPE grid from the left or top, it is
forwarded to neighboring DPEs along the row or column, allowing
multiple DPEs to use the same operand across cycles without repeated
memory access as illustrated in Fig.~\ref{fig:inner1} and
Fig.~\ref{fig:inner2}.

\subsubsection{Inter-Block Locality}

The second type is \textit{inter-block locality}, which emerges from
the diagonal blocking strategy. Since diagonal matrix multiplication
requires each diagonal in $A$ to be multiplied with every diagonal in
$B$, blocking allows partial locality of operand groups across
multiple group-pair computations. A block group from $A$ can be used
across several groupings with different $B$ blocks before it is
evicted from the cache. As shown in Fig.~\ref{fig:outerreuse},
diagonal from $B$(\raisebox{.5pt}{\textcircled{\raisebox{-.9pt} {1}}})
can multiply with all A diagonals inside the cache before it is
evicted. This access pattern improves temporal locality
and reduces DRAM accesses.

\subsubsection{Algorithmic Locality}

\name~also benefits from locality patterns specific to quantum
simulation workloads. In Hamiltonian simulation, matrix exponentiation 
via Taylor expansion, Eq.~\eqref{eq:hamiltonian_simulation}, is commonly used. Repeated sparse
multiplications are performed until convergence. Intermediate results
are written to DRAM, while active portions of matrix~$A$ are reused
from cache. This reuse pattern minimizes unnecessary memory transfers
and improves performance for deep iterative computations.

Together, these three levels of locality, \textit{(1) intra-grid,
(2) inter-block, and (3) algorithmic}, help \name~reduce
memory traffic, lower latency, and sustain high throughput in sparse
quantum workloads. Their combined effect is evaluated in
Section~\ref{sec:eva}.

\begin{figure*}[t]
    \centering

    \begin{subfigure}[t]{0.24\textwidth}
        \centering
        \includegraphics[width=\linewidth]{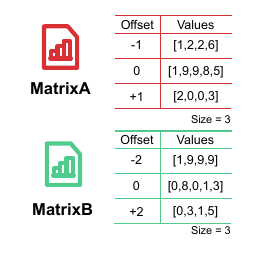}
        \caption{Load Diagonal Matrices.}
        \label{fig:loaddiaq}
    \end{subfigure}
    \hfill
    \begin{subfigure}[t]{0.24\textwidth}
        \centering
        \includegraphics[width=\linewidth]{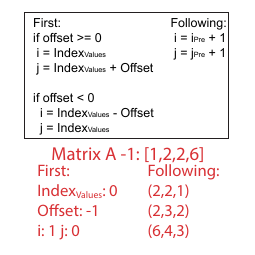}
        \caption{Construct Coordinate Info}
        \label{fig:construct}
    \end{subfigure}
    \hfill
    \begin{subfigure}[t]{0.24\textwidth}
        \centering
        \includegraphics[width=\linewidth]{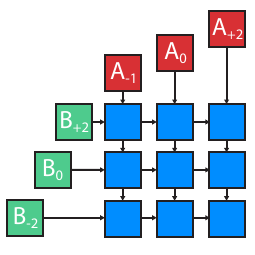}
        \caption{Feed Data}
        \label{fig:feed}
    \end{subfigure}
    \hfill
    \begin{subfigure}[t]{0.24\textwidth}
        \centering
        \includegraphics[width=\linewidth]{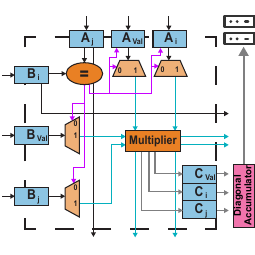}
        \caption{Compute and Accumulate}
        \label{fig:MAC}
    \end{subfigure}

    \caption{Walkthrough Example.}
    \label{fig:walkthrough}
    \vspace*{-0.2in}
\end{figure*}

\subsection{Cycle Analysis}
\label{subsec:cycle}

%
Although \name~inherits the systolic array's pipelined feeding
pattern, its cycle behavior diverges significantly. Dense matrix
multiplication requires $O(N^3)$ cycles, while systolic arrays reduce
this to $O(3N)$~\cite{3N} by exploiting spatial parallelism. Notice
that in highly sparse cases (e.g., $>99\%$ sparsity), dense approaches
waste cycles on zero-valued elements, underutilizing hardware and
memory. \name~overcomes this by applying sparsity-aware scheduling in
diagonal space, executing only valid operations. To model
performance, we decompose \name's cycle count into three phases, (1)
preload, (2) compute, and (3) pop-out, capturing the unique execution
flow of sparse, diagonal-structured workloads.

\subsubsection{Preloading Stage}

Similar to the preloading stage in a traditional systolic array, we
define the end of \name's preload stage as the point when every DPE
has received both of its input operands. The number of cycles required
for this stage is
\begin{equation}
    \text{Cycle}_{\text{Preload}} = R + C - 1,
    \label{eq:pre}
\end{equation}
where $R$ is the number of rows in the DPE grid and $C$ is the number
of columns. $\text{DPE}_{\text{R,C} }$ is last to receive both inputs,
requiring $R-1$ cycles for the top data and $C-1$ cycles for the left
data to propagate. Since the top data is fed at cycle $C$ and the left
data is fed at cycle $R$, the total cycle count for preload is always
$R+C-1$ no matter which arrives first.

\subsubsection{Computation Stage}

We define the computation stage as starting from the cycle immediately
after the preload stage and ending when all input data has been fed
into the DPE grid($\text{T}_{\text{FF} }$). This duration is
\begin{equation}
    \text{Cycle}_{\text{Comp}} = T_{\text{FF}} - \text{Cycle}_{\text{Preload}},
    \label{eq:comp1}
\end{equation}
where $\text{Cycle}_{\text{Preload}}$ is given by Eq.~\eqref{eq:pre} and
$T_{\text{FF}}$ is defined as
\begin{equation}
T_{\text{FF}} =
\begin{cases}
L_{d_{\max}} + R_{d_{\max}}, & d_{\max} \in B,\\
L_{d_{\max}} + C_{d_{\max}}, & d_{\max} \in A.
\end{cases}
\label{eq:ffinish}
\end{equation}
Here, $d_{\max}$ is the index of the longest diagonal, $L_{d_{\max}}$
the length of that diagonal, and $R_{d_{\max}}$ (or $C_{d_{\max}}$)
the row (or column) index at which this diagonal is fed, depending on
whether it originates from $B$ or $A$.


In \name, diagonals are fed in ascending or descending
order. Diagonals near the center are longer, while those near the
edges are shorter. Since the number of diagonals is much smaller than
the matrix size, diagonal lengths decrease at a faster rate than the
simultaneous increase in feed delays. As a result, the overall feed
completion is dominated by the longest diagonal, not the edge of the
array as in traditional systolic designs.

Combining Eq.~\eqref{eq:pre}, \eqref{eq:comp1}, and \eqref{eq:ffinish}, we
derive
\begin{equation}
\text{Cycle}_{\text{Comp}} =
\begin{cases}
L_{d_{\max}} + R_{d_{\max}} - R - C + 1, & d_{\max} \in B, \\
L_{d_{\max}} + C_{d_{\max}} - R - C + 1, & d_{\max} \in A.
\end{cases}
\label{eq:comp}
\end{equation}

This analysis highlights that the computation stage length in \name~is
tightly coupled to the structure of the longest diagonal, underscoring
the importance of diagonal structure in overall cycle efficiency.

\subsubsection{Popout Stage}

The popout stage starts at the cycle after
$T_{\text{FF}}$, when all data has been fed, and ends when the last
partial sum exits the DPE grid at cycle $T_{\text{PF}}$. The cycle
count is
\begin{equation}
    \text{Cycle}_{\text{Popout}} = T_{\text{PF}} - T_{\text{FF}},
    \label{eq:pop1}
\end{equation}
where $T_{\text{FF}}$ is defined in Eq.~\eqref{eq:ffinish}, and $T_{\text{PF}}$ is
\begin{equation}
T_{\text{PF}} =
\begin{cases}
L_{d_{\max}} + R_{d_{\max}} + C - 1 + R - R_{d_{\max}}, & d_{\max} \in B, \\
L_{d_{\max}} + C_{d_{\max}} + R - 1 + C - C_{d_{\max}}, & d_{\max} \in A.
\end{cases}
\label{eq:pfinish}
\end{equation}
Assuming the longest diagonal $d_{\max}$ originates from matrix~B, then
$\text{DPE}_{R,C}$ receives the last $B$ element in its row before it
receives the last A element in its column due to blocking
conditions. $\text{DPE}{R_{d_{\max},C}}$ holds the last A value until
cycle $L_{d_{\max}} + R_{d_{\max}} + C - 1$, at which point it
forwards the data down the column. It then takes $R - R_{d_{\max}}$
additional cycles for this data to reach $\text{DPE}_{R,C}$.

Combining Eq.~\eqref{eq:pop1}, \eqref{eq:ffinish}, and \eqref{eq:pfinish}, we get
\begin{equation}
\text{Cycle}_{\text{Popout}} =
\begin{cases}
R + C - 1 - R_{d_{\max}}, & d_{\max} \in B, \\
R + C - 1 - C_{d_{\max}}, & d_{\max} \in A.
\end{cases}
\label{eq:pop}
\end{equation}

This behavior highlights how the tail latency of the popout phase is
influenced by the relative position of the longest diagonal in the
matrix, further reinforcing the cycle impact of sparse structure in
\name.

\subsubsection{Total Cycle Count}

Combining the results from Eq.~\eqref{eq:pre}, \eqref{eq:comp}, and
\eqref{eq:pop}, the total cycle count for a complete \name{} operation
can be expressed as
\begin{equation}
\text{Cycle}_{\text{Total}} = R + C + L_{d_{\max}} - 1,
\label{eq:cycle}
\end{equation}
where $R$ and $C$ are the number of rows and columns in the DPE grid,
and $L_{d_{\max}}$ is the length of the longest diagonal. Accordingly,
the overall cycle complexity of \name~can be expressed as
\begin{equation}
O\bigl(|D_A| + |D_B| + \max(N_A, N_B)\bigr),
\end{equation}
where $|D_A|$ and $|D_B|$ are the numbers of nonzero diagonals in
matrices $A$ and $B$, respectively, and $N_A$, $N_B$ are the
dimensions of the input matrices.

\noindent\textbf{Remark:} Note that the three
execution stages --- preloading, computation, and popout --- are not
strictly sequential and often overlap in practice. This overlap
manifests when computing individual stage durations, where negative
values may arise due to shared or pipelined cycles across stages. As a
result, the total cycle count provides a more holistic and accurate
measure of execution latency than the isolated stage timings.

\subsection{Walk-Through Example}
\label{subsec:example}

The following steps (corresponding to Fig.~\ref{fig:walkthrough})
depict a walk-through example demonstrating how \textbf{\name}
utilizes diagonal formats to map sparse GEMMs onto DPEs. In this
example, $A$ and $B$ both contain 3 diagonals, and both matrices have
a dimension of $N=5$.  These steps are as follows:

\begin{description}[style=unboxed,leftmargin=0cm]

\item[Step i)] Load two diagonally-compressed matrices to cache.

\item[Step ii)] Construct the matrix metadata .

\item[Step iii)] Instantiate a $3 \times 3$ DPE grid, with 3 diagonals
  from $A$ (columns) and 3 from $B$ (rows). Feed $A$ diagonals
  top-down by ascending index and $B$ left-right by descending index,
  pipelined one cycle apart following systolic scheduling
.

\item[Step iv)] Each DPE receives data with metadata. It compares
  $A$'s $j$ index to $B$'s $i$ index to determine if multiplication
  occurs. Outputs carry $A$'s $i$ and $B$'s $j$ indices and are
  accumulated per diagonal. Once all FIFOs drain, results are written
  back to cache, and the grid is released for the next task.
  
\end{description}

\section{Evaluation}
\label{sec:eva}

\subsection{Methodology}

\subsubsection{Benchmarks}
We select a representative set of matrices from HamLib~\cite{hamlib},
a dataset containing problem Hamiltonians from a variety of
application domains. Specifically, we choose the Transverse Field
Ising Model (TFIM), the Heisenberg model, and several Hubbard models
from the condensed matter domain; the Maximum Cut problem (Max-Cut)
and its quantum variant (Q-Max-Cut) from the binary optimization
domain; and the classical Traveling Salesman Problem (TSP) from the
discrete optimization domain.  The sparsity characteristics of these
HamLib examples are summarized in Table~\ref{tab:dataexample}, and the
associated computational details are provided in
Section~\ref{sec:background}. 
Most Hermitians in the HamLib suite exhibit
diagonal sparsity inherent to their problem definitions and therefore
selected a representative subset from the thousands available.
%
\begin{table}[htb]
\centering
\caption{Representative matrices from HamLib~\cite{hamlib} used in our
  evaluation. Here, \texttt{Dim} denotes the matrix dimensions.
  \texttt{DSparsity} represents the fraction of non-zero diagonals
  relative to all diagonals. \texttt{NNZE} and \texttt{NNZD} indicate
  the number of non-zero elements and non-zero diagonals,
  respectively. \texttt{Iter} refers to the iteration count at which
  the Taylor series converges, as determined by the matrix one-norm.}

\label{tab:dataexample}
\tabcolsep 0pt
\begin{tabular}{@{}P{1.5cm} P{1cm} P{1cm} P{1cm} P{1.5cm} P{1cm} P{1cm} P{1cm}@{}}
\toprule
\textbf{Benchmark} & \textbf{Qubit} & \textbf{Dim} & \textbf{Sparsity}& \textbf{DSparsity} & \textbf{NNZE} & \textbf{NNZD} &\textbf{Iter}\\




\midrule
\multirow{3}{*}{Max-Cut} 
                        & 10 & 1{,}024 & 99.90\% & 99.95\% & 1{,}024 & 1 & 4\\
                        & 12 & 4{,}096 & 99.99\% & 99.99\% & 1{,}936 & 1 & 4\\
                        & 14 & 1{,}6384& 99.99\% & 99.99\% & 16{,}384& 1 & 5\\
                    \midrule
\multirow{3}{*}{Heisenberg} 
                        & 10 & 1{,}024 & 99.46\% & 99.07\%&5{,}632 & 19 & 4\\
                        & 12 & 4{,}096 & 99.84\% & 99.72\%&26{,}624& 23 & 4\\
                        & 14 & 16{,}384& 99.95\% & 99.91\%&122{,}880& 27 & 4\\
                    \midrule
\multirow{2}{*}{TSP}    & 8 & 256      & 99.61\% & 99.80\% &  256   & 1 & 4\\
                        & 15& 32{,}768 & 99.99\% & 99.99\% &32{,}768& 1 & 4\\
\midrule

\multirow{2}{*}{TFIM}                    
                        & 8 & 256      & 96.58\% & 96.67\% & 2{,}240 & 17  & 4\\
                        & 10& 1{,}024  & 98.93\% & 98.97\% & 11{,}264& 21 & 4\\
\midrule
\multirow{2}{*}{Fermi-Hubbard}  
                        &  8& 256      & 98.60\% & 97.46\% &  916    & 13 & 4 \\
                        & 10& 1{,}024  & 99.51\% & 99.17\% & 5{,}120 & 17 & 4\\
\midrule
\multirow{2}{*}{Q-Max-Cut}
                        & 8 & 256      & 98.24\% & 97.06\% & 1{,}152 & 15  & 3\\
                        & 10& 1{,}024  & 99.46\% & 99.07\% & 5{,}632 & 19  & 3\\
\midrule
\multirow{2}{*}{Bose-Hubbard}
                        & 8 & 256      & 99.27\% & 96.28\% & 480     & 19  & 4\\
                        & 10& 1{,}024  & 99.36\% & 98.39\% & 6{,}663 & 33  & 5\\      
\bottomrule
\end{tabular}
\end{table}

\subsubsection{Baseline Accelerators} We compare \name{} against two
state-of-the-art SpMSpM accelerators: SIGMA~\cite{SIGMA} and
Flexagon~\cite{flexagon}. Within Flexagon, we evaluate both the Outer
Product (OP) and Gustavson dataflows. Existing SpMSpM accelerators are limited in
their ability to handle the extreme sparsity patterns found in quantum
simulation. Due to the distinct computational patterns of SpMSpM
compared to sparse-dense matrix multiplication (SpMM) and dense matrix
multiplication (GEMM), we do not compare \name{} against SpMM or GEMM
accelerators such as TPU~\cite{tpu}, Sextans~\cite{song2022sextans},
or Mentor~\cite{lu2024mentor}.
To ensure fair comparison, we standardize the total PE count across
all architectures, setting it equal to the matrix dimension.  While
SIGMA and Flexagon use linear PE layouts with 1-to-1 distribution and
reduction networks, \name{} adopts a 2D DPE grid under the same PE
budget. The grid dimensions are derived from the number of active
diagonals in matrices A and B, prioritizing B due to its static
nature. When diagonal counts exceed feasible hardware limits (e.g.,
10 qubits \textit{Bose-Hubbard} contain 33 non-zero diagonals) we adopt a balanced grid
configuration (e.g., $32 \times 32$) to maintain performance within
the PE constraint(1024 in total). For clarity, all architectures operate under the
same total PE capacity, with \name{} selectively activating only the
subset required by the diagonal structure. In contrast, for
single-diagonal workloads such as \textit{Maxcut} and \textit{TSP}, we
apply a compact $1 \times 4$ pipelined grid to minimize resource usage
while preserving throughput. This flexibility allows \name{} to adapt
efficiently across a wide range of diagonal densities.

\subsubsection{Simulation Infrastructure}
To evaluate the performance of \name{} and baseline designs, we
developed a cycle-accurate model under the STONNE
simulator~\cite{stonne}. Flexagon and SIGMA are both implemented
within the STONNE simulator framework and use the same PE design,
enabling a direct comparison within a unified framework. The 
simulator collects statistics such as the number of multiplications,
FIFO reads/writes, and memory accesses.

For accurate area and power analysis, we implemented the DPE design of
\name{} as well as the PE designs of SIGMA and Flexagon in
Verilog. All processing elements were realized as \texttt{float32}
precision. We synthesized each design using Synopsys' Design Compiler
with a 28nm CMOS standard cell library targeting a clock frequency of
700~MHz.

The power and area estimates, as well as the relative costs of the
\name{} DPE and STONNE PE, are summarized in
Table~\ref{tab:pe_evaluation}. The DPE exhibits a modest area overhead
of $1.05\times$ and a higher power overhead of $1.30\times$ compared
to the STONNE PE under the \texttt{float32} configuration. This
overhead primarily stems from the additional comparator logic and more
sophisticated control structures required to support sparse and
index-aware operations. However, \name{} compensates for this overhead
by significantly reducing total execution cycles. As demonstrated in
Section~\ref{subsec:performance}, this trade-off leads to lower
overall energy consumption for SpMSpM workloads.

\begin{table}[htb]
\centering
\caption{PE Evaluation of \name{} and STONNE.}
\label{tab:pe_evaluation}
\begin{tabular}{@{}lcc@{}}
\toprule
\textbf{Component} & \textbf{Power} (mW) & \textbf{Area ($\mu m^2$)} \\
\midrule
DPE              & 4.3877 (130.77\%)   &   \multirow{5}{*}{7{,}585.20 (105.10\%)}   \\
\multicolumn{1}{l}{\quad-- Multiplier} & 1.6354   &     \\
\multicolumn{1}{l}{\quad-- Comparator} & 0.3247  &      \\
\multicolumn{1}{l}{\quad-- FIFOs} & 0.7568  &      \\
\multicolumn{1}{l}{\quad-- Control \& Others} & 1.6708  &      \\
\midrule
STONNE PE       & 3.3554 (100\%)  &      7{,}214.26 (100\%)\\
\bottomrule
\end{tabular}
 \vspace*{-0.2in}

\end{table}

\subsection{Performance and Energy for SpMSpM}
\label{subsec:performance}

\subsubsection{Performance}
\begin{figure}[t]
    \centering
    \includegraphics[width=\columnwidth]{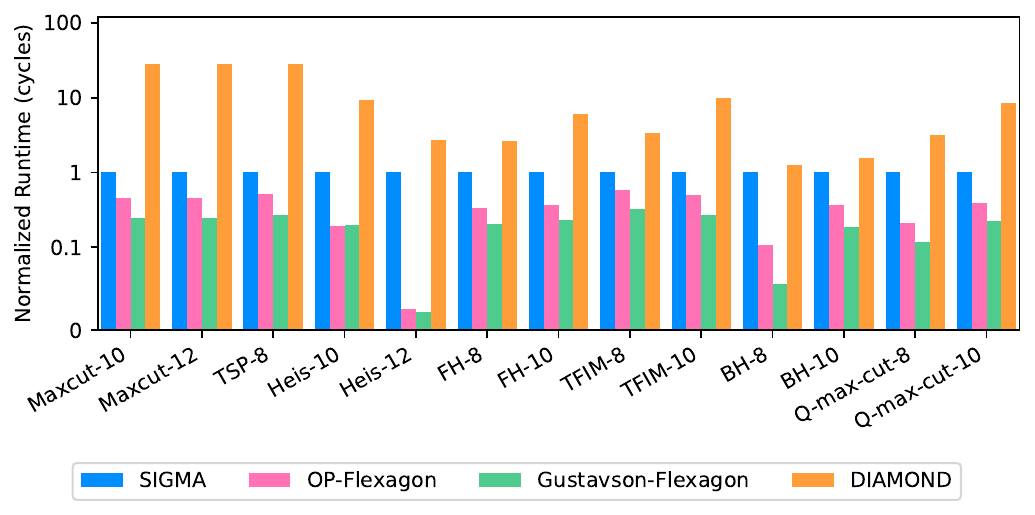}
    \caption{Comparison of performance relative to SIGMA for
      Flexagon and \name{} across seven quantum workloads. Performance is normalized to SIGMA. }
    \label{fig:speedup}
    \vspace*{-0.2in}
\end{figure}

Fig.~\ref{fig:speedup} shows the speedup of our \name{} design over
competing methods across seven quantum benchmarks. In \textit{MaxCut},
\name{} achieves $28\times$, $62\times$, and $113\times$ speedup over
SIGMA, OP, and Gustavson, respectively. For \textit{TSP}, the speedups
are $28\times$, $56\times$, and $106\times$. On the irregular
\textit{Heisenberg} workload, \name{} is $6\times$ faster than SIGMA
and delivers $77\times$–$88\times$ gains over OP and Gustavson. For
\textit{TFIM}, the improvements are $6.7\times$, $13\times$, and
$24\times$.  In \textit{Fermi-Hubbard} and \textit{Q-Max-Cut}, \name{}
runs $4\times$–$6\times$ faster than SIGMA and achieves
$12\times$–$33\times$ speedups over OP and Gustavson. For
\textit{Bose-Hubbard}, \name{} improves performance by $1.4\times$
over SIGMA and $8\times$–$16\times$ over OP and Gustavson. Notably,
\name{} successfully completes all benchmarks, including those with
14+ qubits, while baseline designs fail to finish within the 12-hour
timeout.

These performance advantages stem from \name{}'s diagonal-centric
design, which enables efficient DPE utilization through diagonal
grouping and selective activation. This improves data locality and 
minimizes unnecessary computation. For
instance, in \textit{MaxCut} and \textit{TSP}, where matrices contain
only a single principal diagonal, \name{} computes just the principal
diagonal of matrices, achieving substantial speedups over baselines
that operate on non diagonal formats. As sparsity decreases, such as
in \textit{Heisenberg}, \textit{Fermi-Hubbard}, and
\textit{Q-Max-Cut}, the relative advantage of \name{} diminishes but
remains meaningful due to its ability to align execution with active
diagonal patterns. In contrast, SIGMA incurs substantial overhead from
dense bitmap representations and must allocate large storage
regardless of sparsity.(2 GiB bitmap for TSP-15.) Meanwhile, Outer Product
and Gustavson methods
traverse entire rows or columns, leading to significant inefficiency
when processing diagonal structured sparse matrices. \name{} avoids
these pitfalls by confining computation strictly to nonzero diagonals.

\subsubsection{Energy}
To evaluate energy efficiency, we compare \name{} against SIGMA, 
the most cycle-efficient among the baselines. Flexagon is excluded
from this comparison, as it shares the same simulation backend 
(STONNE) as SIGMA but performs worse than SIGMA on every benchmark.
Including it offers no additional insight and would not affect the
overall energy analysis. Focusing on SIGMA enables a fair and 
representative comparison of per-operation energy cost.

\begin{figure}[t]
    \centering
    \includegraphics[width=\columnwidth]{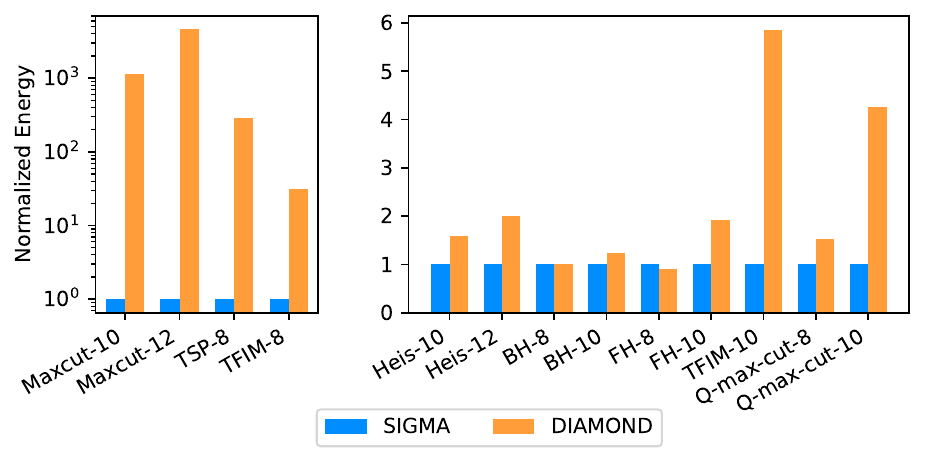}
    \caption{Comparison of energy between SIGMA and
      \name{} across seven quantum workloads. Energy is normalized to SIGMA.}
    \label{fig:energy}
    \vspace*{-0.2in}
\end{figure}

Fig.~\ref{fig:energy} presents energy reductions achieved by \name{}
over SIGMA across various 8, 10, and 12 qubit quantum workloads. For
single-diagonal problems such as \textit{MaxCut-10} and
\textit{TSP-8}, \name{} activates only a minimal $1\times4$ grid (4
for pipelining), in contrast to SIGMA's use of a full $1024$-PE array,
leading to energy savings of $1{,}158\times$ and $290\times$ for the
two benchmarks. At 12 qubits, \name{} continues to outperform SIGMA
with a $4630\times$ saving on \textit{MaxCut-12}. In multi-diagonal
cases that are still sparse in early iterations like \textit{TFIM-10},
\textit{Q-Max-Cut-10}, \textit{Fermi-Hubbard-10},
\textit{Heisenberg-10}, and \textit{Bose-Hubbard-10}, \name{} achieves
savings of $5.86\times$, $4.26\times$, $1.92\times$, $1.59\times$, and
$1.25\times$, respectively. Even when matrix density increases, such
as in the final iterations of TFIM, \name{} maintains its advantage
due to selective DPE activation and reduced data movement in the early
iterations.

These results underscore the strength of \name{}'s diagonal-aware
scheduling: by activating only the necessary DPEs and minimizing
switching activity, memory accesses, and redundant computation,
\name{} delivers substantial energy savings—especially in sparse
scenarios common to quantum simulation.
\begin{figure}[b]
    \vspace*{-0.2in}
  \centering
  \includegraphics[width=\columnwidth]{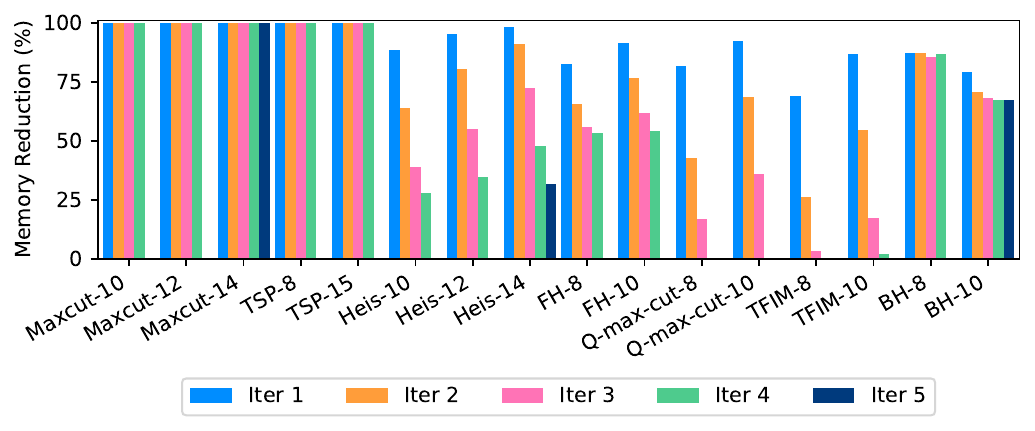}

  \vspace*{-0.1in}
  \caption{Storage Saving on Hamiltonian Simulation.}
  \label{fig:mem}
\end{figure}
\subsection{Memory Footprint}

\subsubsection{Storage Saving}
Fig.~\ref{fig:mem} illustrates the substantial memory savings enabled
by the diagonal storage format, which motivates our \name{} accelerator.
During Hamiltonian simulation, matrix exponentiation is performed
using
Taylor series expansion, where the iteration depth is determined by
the matrix one-norm. As chained multiplications progress, new
diagonals accumulate due to offset additivity, gradually increasing
storage demand. Nevertheless, \name{} preserves high compression in
early iterations. In \textit{MaxCut} and \textit{TSP}, where only the
principal diagonal is active, memory savings remain above 99\%
throughout. Even in denser workloads like \textit{Heisenberg},
\textit{Fermi-Hubbard}, and \textit{Q-Max-Cut}, \name{} achieves 60\%–98\%
savings in early steps and retains 31\%–48\% at
convergence. \textit{Bose-Hubbard} and \textit{TFIM} also see 67\%–87\%
early savings, though benefits taper off as the diagonal count
increases. Crucially, by storing only active diagonals, \name{}
operates independently of full matrix dimensions, enabling
it to handle much larger matrices than conventional accelerators with
fixed-size dense buffers or row/column-based sparsity. This
diagonal-aware compression is key to both memory efficiency and
scalability across quantum workloads.

\subsubsection{Data Cache Locality}
\begin{figure}[t]
    \centering
    \includegraphics[width=\columnwidth]{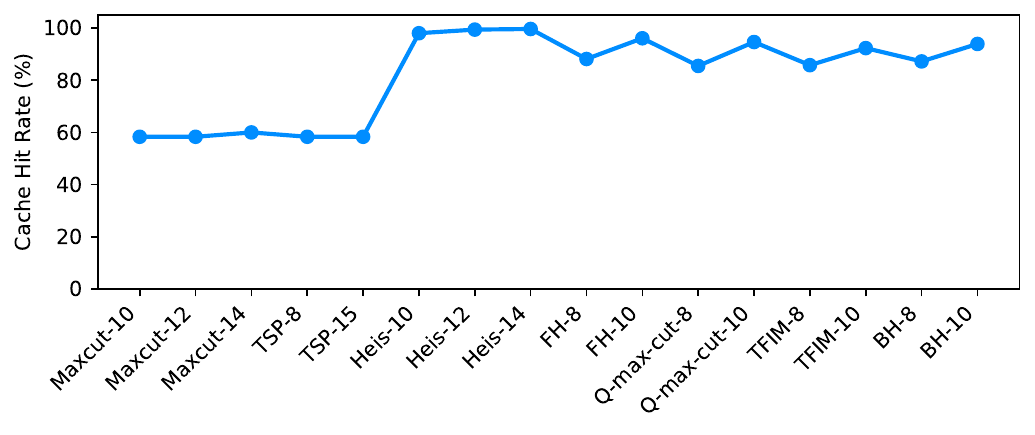}
    \caption{Cache hit rate using a 2-set, 2-way cache.}
    \label{fig:cache}
    \vspace*{-0.2in}
\end{figure}

After applying the blocking strategy in \name{}, cache hit rates 
exceed 90\% across all multi-diagonal benchmarks, as shown in Fig.
~\ref{fig:cache}. Denser workloads such as \textit{Heisenberg-10/12/14}
(hit rates of 98.0\%, 99.4\%, 99.6\%), \textit{Fermi-Hubbard-10} 
(96.1\%), \textit{TFIM-10} (92.3\%), \textit{Bose-Hubbard-10} (93.9\%), and 
\textit{Q-Max-Cut-10} (94.6\%) benefit substantially from blocking. 
This strategy maps each diagonal group to a dedicated cache line 
and completes all operations before switching, thereby maximizing 
temporal locality and minimizing evictions.

In contrast, \textit{MaxCut} and \textit{TSP} benchmarks contain
only a single diagonal; blocking has no effect, and their hit rates
remain around 58.3\%, reflecting only compulsory misses. Overall,
the results validate that blocking is particularly effective for
workloads with numerous diagonals and dense reuse, significantly 
boosting cache efficiency in \name{}.

\subsection{Analysis}

\name{} excels when nonzeros are concentrated in a few dense diagonals
near the principal axis, enabling compact mapping to the DPE grid with
high data reuse, minimal routing, and consistent utilization, even as
matrix dimensions grow. Unlike traditional architectures that scale
with matrix size, \name{} decouples performance from overall
dimensionality by operating only on active diagonals. However,
performance diminishes when nonzeros spread across many distant
diagonals, particularly near matrix corners, which leads to fragmented
compute regions, lower reuse, and increased idle cycles. In
Hamiltonian simulation, the diagonal count tends to grow due to offset
additivity, as shown in Fig.~\ref{fig:mem}, 
but early iterations typically remain compact. These early
stages dominate both runtime and energy, allowing \name{} to maintain
efficiency before diagonal spread becomes a limiting factor.

\section{Related Work}
\label{sec:related_work}

SpMSpM is a fundamental primitive in machine learning, graph
analytics, and scientific computing. Several accelerators have been
proposed to address irregular sparsity. Early outer product designs
such as OuterSPACE~\cite{outerspace} partition output matrices into
partial products but incur heavy off-chip traffic.
SpArch~\cite{sparch} scales better with high-radix mergers and compact
representations, yet complex mergers struggle at extreme
sparsity. ExTensor~\cite{extensor} uses a hybrid outer/inner dataflow
for sparse tensor algebra but, like its predecessors, fails to exploit
structured sparsity.

Inner-product designs such as SIGMA~\cite{SIGMA} maintain compute
utilization with bitmap intersections but degrade on high-dimensional
matrices. Row-wise designs based on Gustavson’s algorithm, e.g.,
Gamma~\cite{gamma}, reduce off-chip traffic by accumulating full
rows. Flexagon~\cite{flexagon} adds reconfigurability across
inner/outer/Gustavson dataflows but still treats diagonal matrices
like any other sparse operand. CSR-based accelerators such as
MatRaptor~\cite{matraptor} similarly target general sparsity; at high
sparsity, compressed formats lose storage benefit and induce irregular
memory access, reducing arithmetic intensity and stalling SIMD
pipelines.

Architectures such as FEASTA~\cite{feasta} extend beyond SpMSpM to
general sparse tensor algebra through iterative fiber joins, but their
flexibility sacrifices specialization and, like the above designs,
ignores diagonal structure. Accelerators for Sparse-Dense matrix
multiplication (SpMM) or dense matrices, e.g., TPU~\cite{tpu},
Sextans~\cite{song2022sextans}, and Mentor~\cite{lu2024mentor}, target
different computational patterns. Likewise, high-performance quantum
simulators and FPGA-based emulators focus on gates or state-vectors,
not sparse matrix kernels in Hamiltonian evolution.

To our knowledge, no prior work optimizes for the diagonal sparsity
common in Hermitian matrices of Hamiltonian-based quantum
simulations. Our architecture, \name{}, is the first SpMSpM accelerator
to be diagonal-aware. By transforming diagonal and near-diagonal
matrices into dense systolic-array computations, \name{} removes costly
fiber intersections and merges, significantly improving utilization and
performance. Existing accelerators treat all sparsity uniformly and thus
cannot exploit structured diagonality.
\section{Conclusion}
\label{sec:conclusion}

In this paper, we proposed \name{}, a co-designed systolic accelerator
for SpMSpM operating in diagonal space. By feeding only the nonzero
diagonals in diagonal-oriented formats, \name{} decouples performance
from matrix dimensions and scales with diagonal sparsity. Its Diagonal
Processing Elements (DPEs) support index matching, data holding, data
forwarding, and multiplication, enabling efficient execution of
structured quantum workloads. Evaluated on benchmarks from HamLib,
\name{} achieves average performance improvements of $10.26\times$,
$33.58\times$, and $53.15\times$ over SIGMA, Outer Product, and
Gustavson-style baselines (the latter two from Flexagon) with peak
speedups up to $127.03\times$ while reducing energy consumption by an
average of $471.55\times$ and up to $4630.58\times$ compared to
SIGMA. These results demonstrate that diagonal-centric reformulation,
combined with hardware-aware co-design, which provides a scalable and
energy-efficient solution for sparse matrix computation in quantum
simulation and beyond.



\bibliographystyle{IEEEtranS}
\bibliography{refs}

\end{document}